\documentclass[preprint]{elsarticle}
\usepackage[utf8]{inputenc}
\usepackage{natbib}
\usepackage{amsmath}
\usepackage{amsfonts}
\usepackage{amssymb}
\usepackage{graphicx}
\usepackage{tabularx}
\usepackage{booktabs}
\usepackage{siunitx}
\usepackage{wrapfig}
\usepackage[version=3]{mhchem}
\usepackage{subfigure}
\usepackage{caption}
\usepackage{xcolor}
\biboptions{sort&compress}	
\usepackage{fancyhdr}
\newenvironment{keywords}%
   {\begin{trivlist}\item[]{\bfseries\sffamily Keywords:}\ }
   {\end{trivlist}}

\journal{Elsvier, License CC-BY-NC-ND 4.0}
\title{RF Helicon-based Inductive Plasma Thruster (IPT) Design for an Atmosphere-Breathing Electric Propulsion System (ABEP)}

\author[irs]{F.~Romano\corref{cor1}\fnref{fn1}}
\ead{romano@irs.uni-stuttgart.de}

\author[irs]{Y.-A.~Chan\fnref{fn2}}
\ead{binder@irs.uni-stuttgart.de}

\author[irs]{G.~Herdrich\fnref{fn3}}
\ead{herdrich@irs.uni-stuttgart.de}

\author[irs]{C.~Traub}
\author[irs]{S.~Fasoulas}
\author[UniMAN]{P.C.E.~Roberts}
\author[UniMAN]{K.~Smith}
\author[UniMAN]{S.~Edmondson}
\author[UniMAN]{S.~Haigh}
\author[UniMAN]{N.H.~Crisp}
\author[UniMAN]{V.T.~A.~Oiko}
\author[UniMAN]{S.D.~Worrall}
\author[UniMAN]{S.~Livadiotti}
\author[UniMAN]{C.~Huyton}
\author[UniMAN]{L.A.~Sinpetru}
\author[UniMAN]{A.~Straker}

\author[Deimos]{J.~Becedas}
\author[Deimos]{R.M.~Domínguez}
\author[Deimos]{D.~González}
\author[Deimos]{V.~Cañas}
\author[Deimos]{V.~Sulliotti-Linner}

\author[GomSpace]{V.~Hanessian}
\author[GomSpace]{A.~Mølgaard}
\author[GomSpace]{J.~Nielsen}
\author[GomSpace]{M.~Bisgaard}

\author[UPC]{D.~Garcia-Almiñana}
\author[UPC]{S.~Rodriguez-Donaire}
\author[UPC]{M.~Sureda}

\author[UCL]{D.~Kataria}

\author[CNU]{R.~Outlaw}

\author[Eurocons]{R.~Villain}
\author[Eurocons]{J.S.~Perez}
\author[Eurocons]{A.~Conte}
\author[Eurocons]{B.~Belkouchi}

\author[concentris]{A.~Schwalber}
\author[concentris]{B.~Heißerer}

\fntext[fn1]{Research Associate, romano@irs.uni-stuttgart.de}
\fntext[fn2]{Research Associate, chan@irs.uni-stuttgart.de}
\fntext[fn3]{Head Plasma Wind Tunnels and Electric Propulsion, herdrich@irs.uni-stuttgart.de}

\address[irs]{Institute of Space Systems (IRS), University of Stuttgart, Stuttgart, 70569, Germany}
\address[UniMAN]{The University of Manchester, George Begg Building, Sackville Street, Manchester, M13 9PL, UK}
\address[Deimos]{Elecnor Deimos Satellite Systems, C/ Francia 9, 13500, Puertollano, Spain}
\address[GomSpace]{GomSpace AS, Langagervej 6, Aalborg East 9220, Denmark}
\address[UPC]{UPC-BarcelonaTECH, Colom 11, TR5 - 08222 Terrassa, Spain}
\address[UCL]{Mullard Space Science Laboratory, University College London Holmbury St. Mary, Dorking, Surrey, RH5 6NT, UK}
\address[CNU]{Christopher Newport University, Newport News, Virginia 23606, United States}
\address[Eurocons]{Euroconsult, 86 Boulevard de Sébastopol, Paris, France}
\address[concentris]{concentris research management gmbh,Ludwigstraße 4, Fürstenfeldbruck, 82256, Germany}
\begin{document}

\begin{abstract}
Challenging space missions include those at very low altitudes, where the atmosphere is source of aerodynamic drag on the spacecraft. To extend such missions lifetime, an efficient propulsion system is required. One solution is Atmosphere-Breathing Electric Propulsion (ABEP). It collects atmospheric particles to be used as propellant for an electric thruster. The system would minimize the requirement of limited propellant availability and can also be applied to any planet with atmosphere, enabling new mission at low altitude ranges for longer times. Challenging is also the presence of reactive chemical species, such as atomic oxygen in Earth orbit. Such species cause erosion of (not only) propulsion system components, i.e.~acceleration grids, electrodes, and discharge channels of conventional EP systems. IRS is developing within the DISCOVERER project, an intake and a thruster for an ABEP system. The paper describes the design and implementation of the RF helicon-based inductive plasma thruster (IPT). This paper deals in particular with the design and implementation of a novel antenna called the birdcage antenna, a device well known in magnetic resonance imaging (MRI), and also lately employed for helicon-wave based plasma sources in fusion research. This is aided by the numerical tool XFdtd\textsuperscript{\textregistered}. The IPT is based on RF electrodeless operation aided by an externally applied static magnetic field. The IPT is composed by an antenna, a discharge channel, a movable injector, and a solenoid. By changing the operational parameters along with the novel antenna design, the aim is to minimize losses in the RF circuit, and accelerate a quasi-neutral plasma plume. This is also to be aided by the formation of helicon waves within the plasma that are to improve the overall efficiency and achieve higher exhaust velocities. Finally, the designed IPT with a particular focus on the birdcage antenna design procedure is presented. 

\end{abstract}
 \maketitle
\begin{keywords}
ABEP - IPT - VLEO - Helicon - Birdcage
\end{keywords}

\section*{Nomenclature}
\noindent
ABEP: Atmosphere-Breathing Electric Propulsion\\
IPT: RF Helicon-based Inductive Plasma Thruster\\
VLEO: Very Low Earth Orbit\\
S/C: Spacecraft 
\section{Introduction} 
This work presents the RF helicon-based inductive plasma thruster (IPT) designed at the Institute of Space Systems (IRS) of the University of Stuttgart within the DISCOVERER project, that aims to redesign very low Earth orbit (VLEO), for altitudes $h<\SI{400}{\kilo\meter}$~\cite{Crisp2020}, platforms by researching on low drag materials, aerodynamic attitude control systems, and by the development of an Atmosphere-Breathing Electric Propulsion system (ABEP)~\cite{romanoiepc,romanosp2016,romanoacta,romanoiepc2,romanosp2018,romanoiepc3,romanoiac3} for which the IPT is designed. 
\begin{figure}[h]
	\centering
	\includegraphics[width=12.5cm]{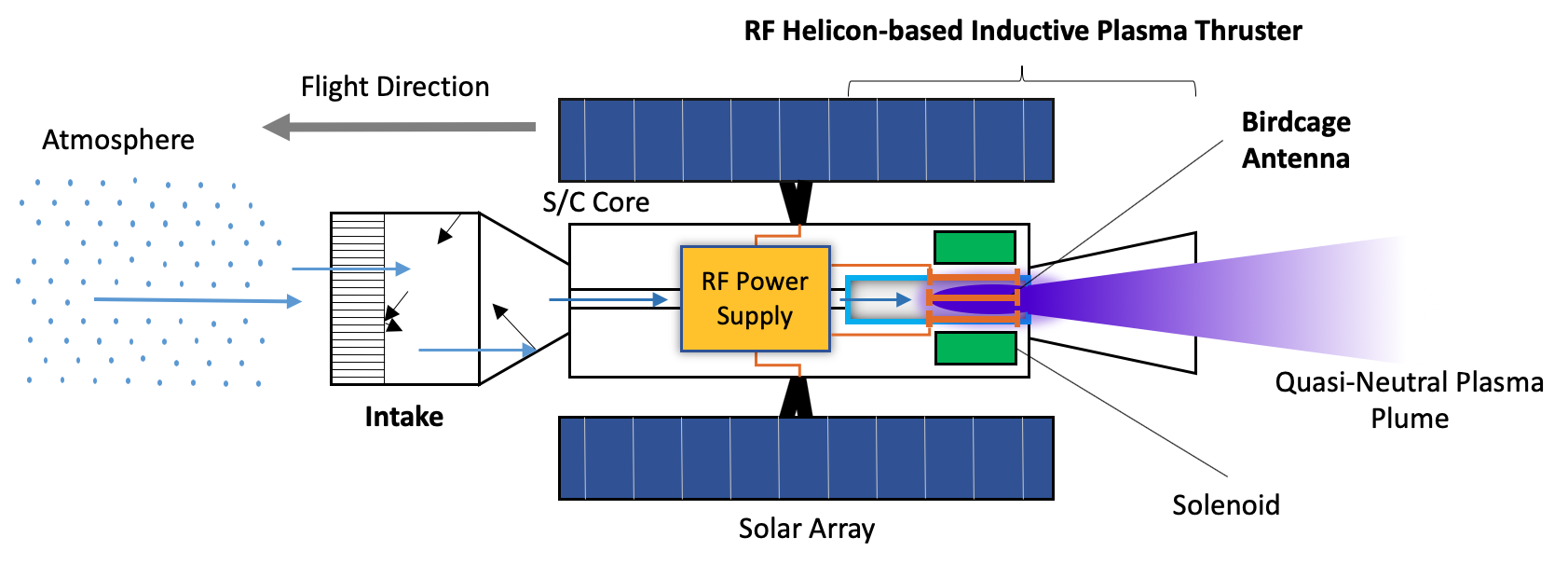}
	\caption{Atmosphere-Breathing Electric Propulsion Concept using an IPT.}
	\label{fig:ABEP}
\end{figure}
An ABEP system, see Fig.~\ref{fig:ABEP}, collects atmospheric particles by means of an intake and uses them as propellant for an electric thruster. The system is theoretically applicable to any planet with an atmosphere, and will drastically reduce the requirement of on-board propellant storage while extending the mission's lifetime. The main atmospheric species in VLEO are \ce{N2} and \ce{O}~\cite{romanoacta}. Atomic oxygen, especially, will rapidly degrade thruster components that are in direct contact with it. Precedent studies showed such degradation phenomena when operating conventional EP with atmospheric propellant, especially on accelerating grids in RF ion thrusters (RIT), and discharge channels in Hall-effect thrusters (HET)~\cite{presitael1,presitael2}. While many ABEP concepts have been investigated in the past based on RIT~\cite{di2007ram,presitael1,presitael2}, HET~\cite{busek,busek2,SITAEL2016,SITAEL2017,SITAEL2019a,SITAEL2019b}, and RF plasma thruster~\cite{shabshelowitz2013study}, the only laboratory tested ABEP system to date is the RAM-HET that includes intake and HET into one device~\cite{SITAEL2016,SITAEL2017,SITAEL2019a,SITAEL2019b}. Results shows that the thruster unit needs further optimization to operate on atmospheric propellant as the thrust generated is still too low compared to the expected drag. In particular, the RAM-HET also requires injection of \ce{Xe} to aid the ignition of the thruster, as well as to operate the neutraliser, all in all implying the need of an extra propellant tank and respective feeding system or, at least, the need of a neutraliser capable of long-time operation on atmospheric propellant~\cite{SITAEL2016,SITAEL2017,SITAEL2019a,SITAEL2019b}. Finally, the IPT design aims to solve these issues by:
\begin{itemize}
\item Employing RF contactless technology for ionization and acceleration;
\item Minimize power losses by providing electrically optimized antenna operating at resonance condition;
\item Accelerating both ions and electrons at the same time for a quasi-neutral plasma exhaust that does not require a neutraliser.
\end{itemize} 
IPT belongs to the category of electrodeless RF plasma thrusters, devices in which an antenna fed by RF power is located around the discharge channel where the propellant is injected, ensuring no direct contact with it. The antenna generates the electromagnetic (EM) fields that ionize the propellant that is then ejected at high velocity to generate thrust. Particularly, the IPT belongs to the subcategory of the helicon plasma thrusters. Those devices are already studied in literature but generally use different kind of antennae, mainly half-helical, Nagoya type, or coil type~\cite{Melazzi_2015,isayama2018review}. The RF discharge is aided by an external static magnetic field applied axially that provides the necessary conditions for the formation of helicon waves within the plasma, delivering increased plasma density and better power coupling efficiency~\cite{chen224,EPFL4}. The currently investigated helicon plasma thrusters are those from the Australian National University (ANU)~\cite{charles2006helicon}, the University of Madrid, Spain~\cite{ahedo2019helicon},  the University of Maryland, USA~\cite{vitucci2019development}, the company T4i, Italy~\cite{manente2019regulus}, the Tohoku University, Japan~\cite{takahashi2019helicon}, the Tokyo University of Agriculture and Technology, Japan~\cite{isayama2018review}, and the Washington University, USA~\cite{vereen2019recent}. The company Ad Astra also includes an helicon plasma stage for the VASIMR thruster~\cite{squire2019steady}. The maximum helicon plasma thruster thrust efficiency as of 2019 reached $\eta_T\sim 20\%$~\cite{takahashi2019helicon}. 

As anticipated in Fig.~\ref{fig:ABEP}, the IPT in the ABEP system is based on a cylindrical birdcage antenna. Such antenna has been investigated for space propulsion in~\cite{EPFL_helicon_2005}, but it is the first time that it is implemented in a plasma thruster. Currently, birdcage antennae are used at EPFL as plasma source for fusion research, called RAID, in which helicon waves have been measured~\cite{EPFL1,EPFL2,EPFL3,EPFL4}. The birdcage antenna reduces overall power consumption and matching requirements by operating at resonance condition, and it is designed to provide an EM field configuration that both ionizes the propellant and is expected to contribute to the acceleration of a quasi-neutral plasma plume. The static magnetic field generated by the solenoid provides the boundary conditions required for the formation of helicon waves. This paper describes the IPT concept and design supported by the theoretical description of the birdcage antenna, its design procedure, its theoretical interaction with the plasma as well as the implementation in the IPT and its preliminary verification. 

\section{RF helicon-based IPT Design}
The starting point of the IPT development has been the heritage of IRS on inductive plasma generators (IPG)~\cite{georg1,georg2,georg16}. Previous work on ABEP has been performed by using the small scale IPG6-S in the last years as test-bed for the ABEP thruster, such device is based on a coil-type antenna fed at \SI{4}{\mega\hertz} for up to \SI{3.5}{\kilo\watt} input power, but not optimized for propulsion purposes,~\cite{romanoacta,romanoiepc2}. Recent test campaigns have shown a large increase of power absorption by applying an external magnetic field to IPG6-S~\cite{romanorgcep2}. This confirmed the feasibility of an improved plasma source that makes use of an applied static magnetic field, especially as it creates the required boundary conditions to trigger excitation of helicon waves~\cite{chen224}.

\subsection{IPT Concept}
\begin{figure}[h]
	\centering
	\includegraphics[width=12cm]{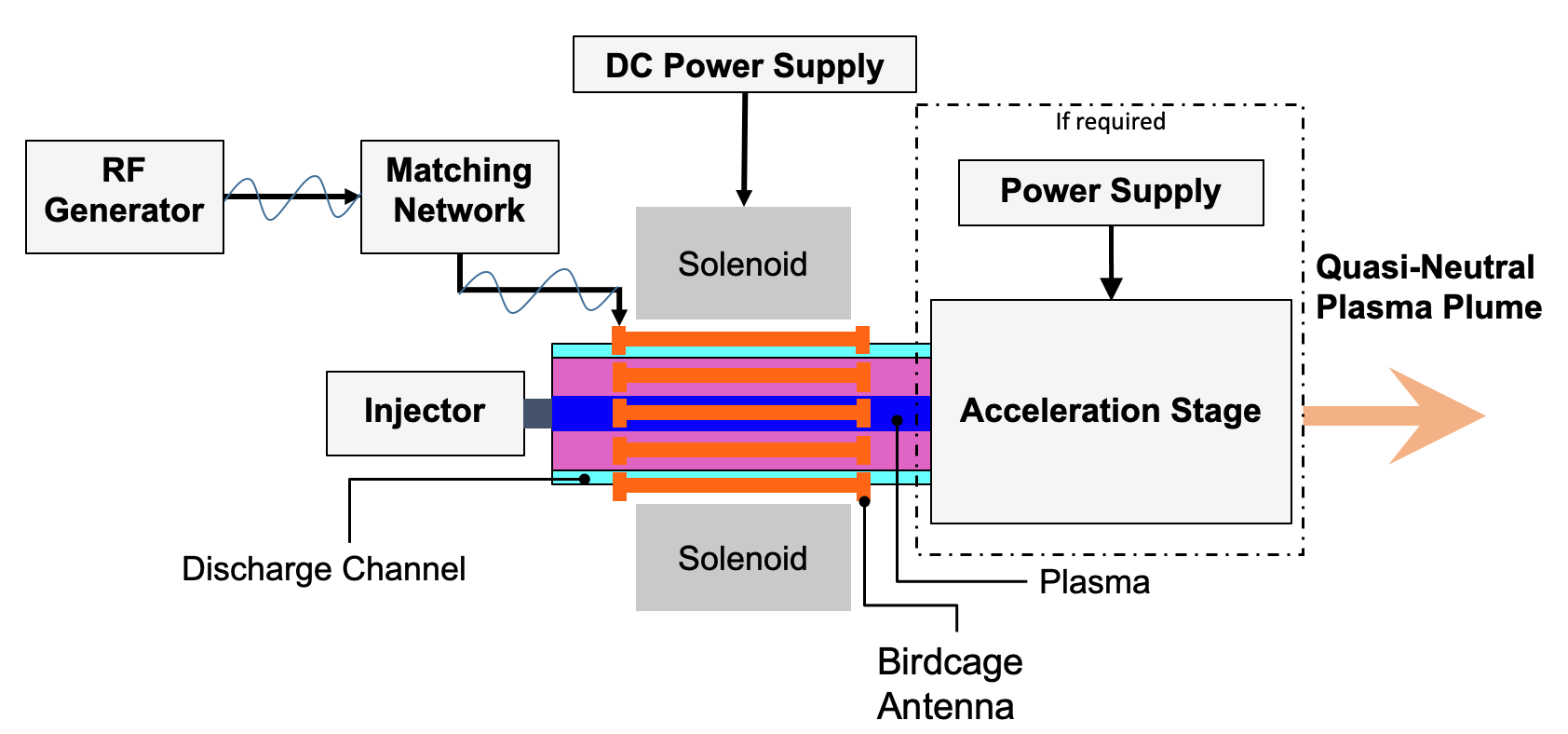}
	\caption{RF Helicon-based Inductive Plasma Thruster (IPT) Concept.}
	\label{fig:IPT}
\end{figure}

The IPT is based on an RF-fed cylindrical birdcage antenna combined with an externally applied static magnetic field for the ionization of the propellant, the triggering of helicon waves in the plasma, and the acceleration and ejection of a quasi-neutral plasma plume. The concept is shown in Fig.~\ref{fig:IPT}. An RF generator combined with a matching network provides the RF power to the birdcage antenna. Propellant is injected by the injector/intake into the discharge channel that is surrounded by the birdcage antenna. An external static magnetic field, generated by a solenoid fed by a DC power supply, is applied along the axis of the discharge channel. The EM fields generated by the birdcage antenna ionize the propellant and provide, in conjunction with the applied static magnetic field, the respective acceleration. This provides contactless operation of the thruster and the acceleration of a quasi-neutral plasma plume while providing a partially matched load for the RF generator, therefore reducing overall matching requirements, therefore power losses.

\subsection{IPT Laboratory Model Set-Up}
\begin{figure}[h]
	\centering
	\includegraphics[width=7cm]{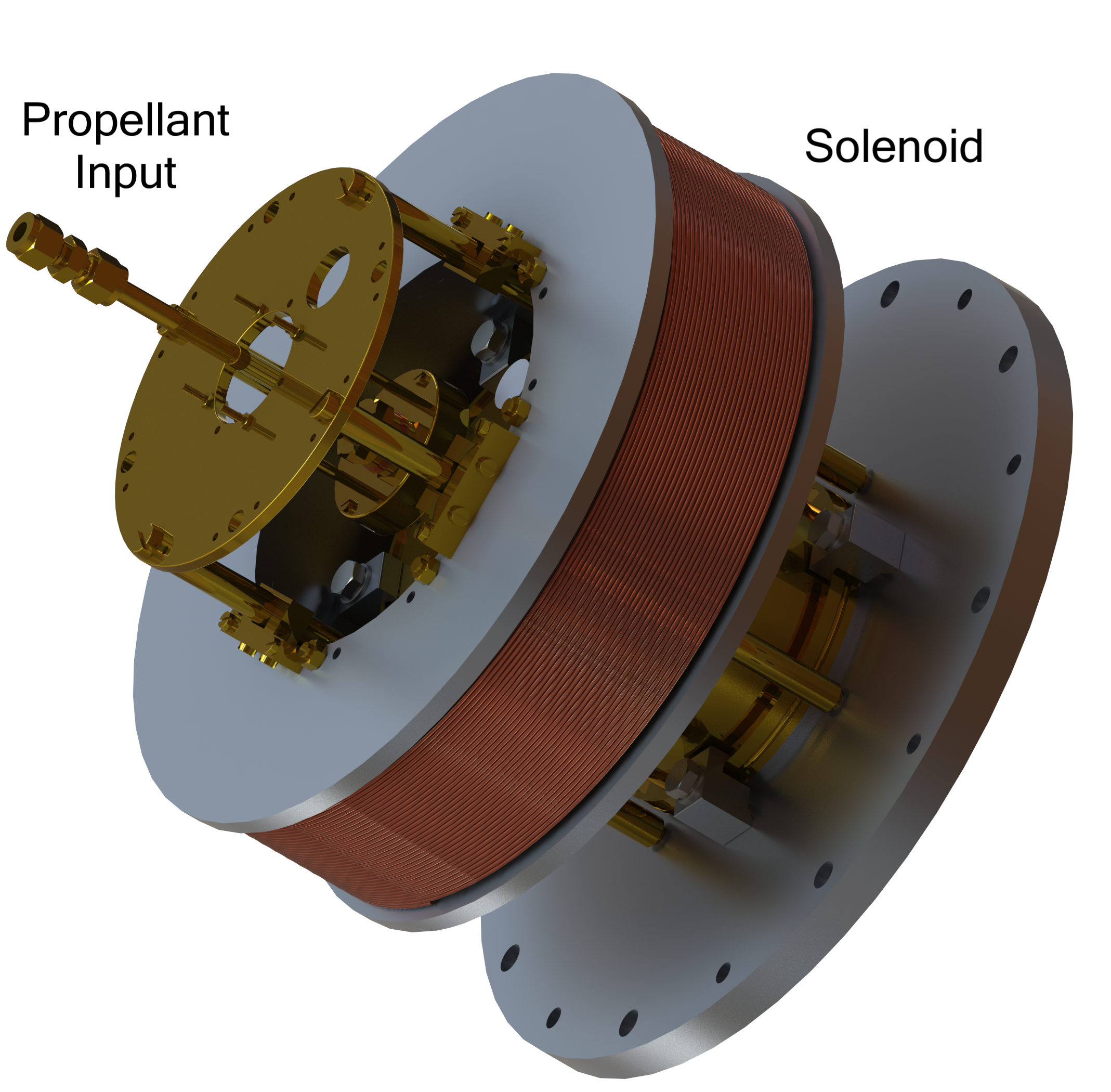}
	\caption{IPT Rendering with Solenoid.}
	\label{fig:IPTrender}
\end{figure}

\begin{figure}[h]
	\centering
	\includegraphics[width=9cm,angle=180]{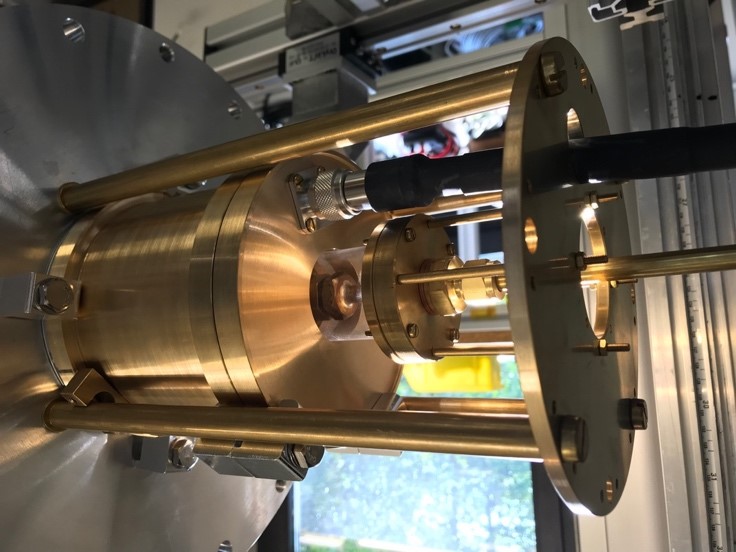}
	\caption{IPT Assembled (Solenoid not integrated)~\cite{romanoiepc3}.}
	\label{fig:assembledIPT}
\end{figure}

The IPT is a laboratory model designed for maximum (technical) flexibility and passive cooling, to allow easy modifications for optimization purposes, and it is composed of: the propellant injector, the discharge channel, the birdcage antenna, the shield, the solenoid, and the supporting structure, see Fig.~\ref{fig:IPTrender}.
The injector is movable along the symmetry axis $z$, as well as the external solenoid, that can produce a magnetic field up to \SI{70}{\milli\tesla} with \SI{15}{\ampere} DC current. At such current, more than \SI{30}{\minute} operation is possible without overheating. This to allow plasma diagnostic measurements to be performed. The IPT structure is made of brass to minimize Eddie currents due to the RF fields, and to minimize interactions with the externally applied static magnetic field. Enclosed within a brass RF shield is the birdcage antenna. The shield isolates the outer environment from the EM fields created by the birdcage and vice versa. Mechanical design is such to allow mounting on standard ISO-K flanges therefore minimizing the in-house required parts and enabling testing on standard facilities. The RF power is provided by an RF-Generator connected to an auto-matching network. The integrated IPT, without solenoid, is shown in Fig.~\ref{fig:assembledIPT}.

\subsection{IPT Antenna Design Tools}
The software used for the birdcage antenna design is the commercial software Remcom Inc. XFdtd\textsuperscript{\textregistered} 7.8.1.3. It is a full wave 3D EM simulation software based on the finite-difference time domain (FDTD) method~\cite{xfdtd}, especially developed for antennae design. It enables antennae modelling and EM field simulation of complex devices. The main advantage is that it has built-in functions for birdcage antennae design, moreover, the full 3D IPT CAD can be loaded into it, thus optimizing the simulation set-up and reducing uncertainties due to e.g.~conductive components placed near the antenna and, finally, it can visualize the resulting 3D EM fields. To experimentally verify the birdcage antenna design, the portable spectrum analyzer Rohde \& Schwarz FSH4 is used to record the frequency response and to qualitatively compare it against numerical results, it is a portable device that provides real time antenna performance and, therefore, can be used to perform the manual fine tuning of the antenna, as well as to relatively quickly check the tuning before each test.

\section{IPT RF Circuit Optimization, and the Birdcage Antenna}
The birdcage antenna of the IPT, is the RF-fed component that creates the EM fields required for the ionization and acceleration of the propellant. Any antenna requires to be tuned at the correct frequency and matched to the required impedance in order to provide maximum power coupling. This section presents the RF circuit requirements and the respective design choices, the birdcage antenna theory of operation, the choice of the resonance mode for operation, and the resulting EM fields.

\subsection{IPT RF Circuit}
Most RF generators deliver the power at a source impedance of $Z_S=50+j0\SI{}{\ohm}$. The power transfer from a source (RF generator) to its load (antenna and plasma) is maximized only if the load’s impedance $Z_L$ is matched to that of the source $Z_S$. Therefore the target impedance of the antenna is fixed to the given output of the available generator. In terms of operating frequency, this can be first chosen depending on the application, and then tuned.

Literature~\cite{Chen_2015} and preliminary numerical results presented in~\cite{romanorgcep2} with both HELIC~\cite{HELIC_Theory_I,HELIC_Theory_II} and ADAMANT~\cite{ADAMANT} numeric tools,  lead to a required frequency equal or higher than \SI{27.12}{\mega\hertz}, as it leads to easier ignition at low pressure, and better power absorption at higher plasma densities~\cite{chen233}. Finally, a \SI{4}{\kilo\watt}, \SI{40.68}{\mega\hertz} power supply with an auto-matching network has been acquired, therefore determining the IPT operating frequency. 

\begin{figure}[h]
	\centering
	\includegraphics[width=11cm]{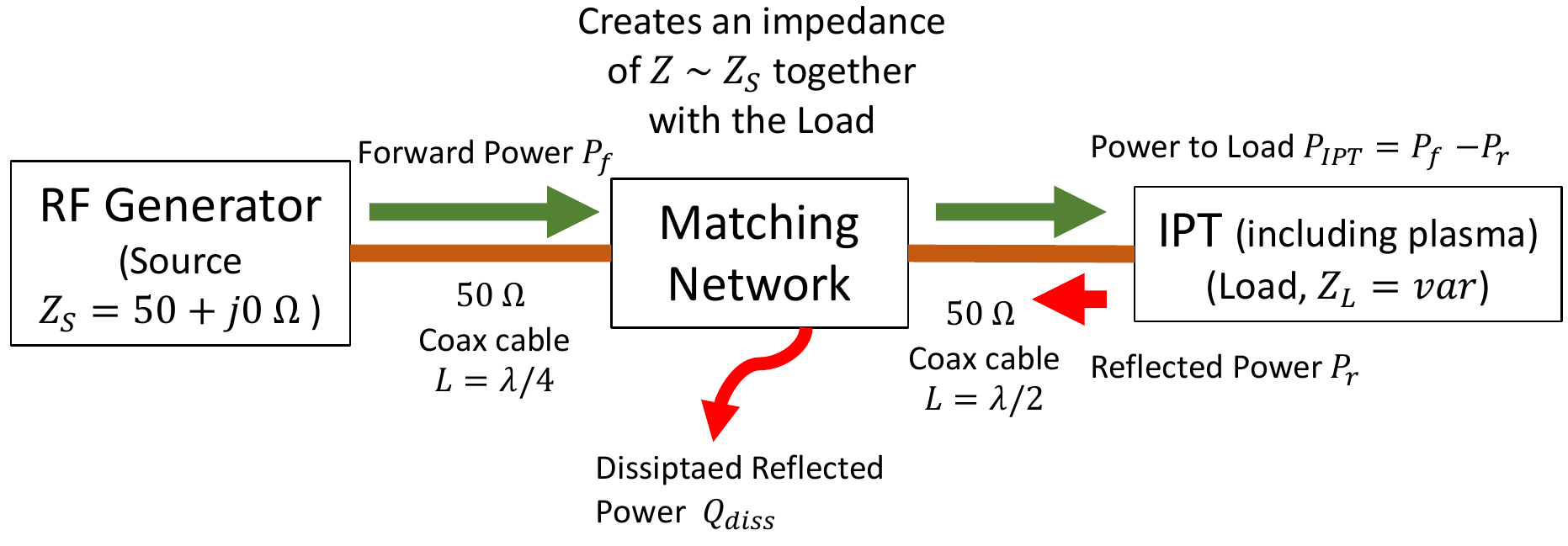}
	\caption{Simplified IPT RF Circuit.}
	\label{fig:imped}
\end{figure}

Not only the antenna is to be considered in the IPT design, but also the whole RF circuit: RF generator, matching network, connectors, and cables~\cite{walkerRF}. The matching network is introduced to create a resonant circuit with the load that is dynamically matched to $Z_S$ by a system of variable inductors and capacitors, so that the forward power $P_f$ is completely absorbed between the IPT, that absorbs $P_{IPT}$, and the matching network absorbing $P_r$, this being the reflect power from the IPT, see Fig.~\ref{fig:imped}. The matching network works mainly as protection for the RF generator, but it does not improve the load itself. Therefore, an optimum design of the antenna is required to maximize the power transfer from source to load. Plasma is finally a variable impedance that implies the need of a dynamic matching control, therefore a matching network, at least for laboratory purposes, is quasi-mandatory to protect the RF generator. An accurate selection of cabling, connectors and, finally, of the antenna has been performed. The schematics describing the simplified RF circuit of the IPT including respective RF generator, matching network and connecting cables, is shown in Fig.~\ref{fig:imped}.

Particularly, the coaxial cable connecting RF generator to matching network has been chosen with a length of $L=\lambda/4$~$\SI{50}{\ohm}$, where $\lambda$ is the RF wavelength corrected by the velocity factor $VF$ of the coaxial cable, and the one connecting matching network and IPT with a length of $L=\lambda/2$~$\SI{50}{\ohm}$ both using an~$RG393/U$ coaxial cable. As in general, the impedance can be transformed by choosing different lengths of coaxial cables, the choice of $L=\lambda/2$ ensures not to transform $Z_{L}$, therefore the matching network matches directly $Z_{L}$, leading to reduced losses due to mismatches, but also reducing misreadings of forward and reflected powers~\cite{walkerRF}.

In general, the antenna (and the plasma), are treated as an impedance $Z$, see Eq.~\ref{eq:impedance}, with a both real, resistance, and imaginary, reactance, parts~\cite{ADAMANT}. In particular, the reactance $X$ is the sum of an inductive component, characterised by its inductance $L$, and a capacitive one, characterised by a capacitance $C$, both directly linked to the operating frequency. In general, the overall reactance has to minimized to reduce power losses in RF circuits. The acquired power supply operating at $f=\SI{40.68}{\mega\hertz}$ leads to an "a priori" higher reactance to the RF circuit, as it is partly directly proportional to the applied frequency, compared to common RF plasma sources/thruster operating at $f=\SI{13.56}{\mega\hertz}$, and has to be zeroed for the matching.

\begin{equation}
\begin{aligned}
\vec{Z} &=\vec{R}+j\vec{X}\\
X & =X_L+X_C=2\pi fL+\biggl(\frac{1}{2\pi fC}\biggr)
\end{aligned}
\label{eq:impedance}	
\end{equation}

\subsection{Theoretical Description of the Birdcage Antenna}
\begin{figure}[h]
	\centering
	\includegraphics[width=10cm]{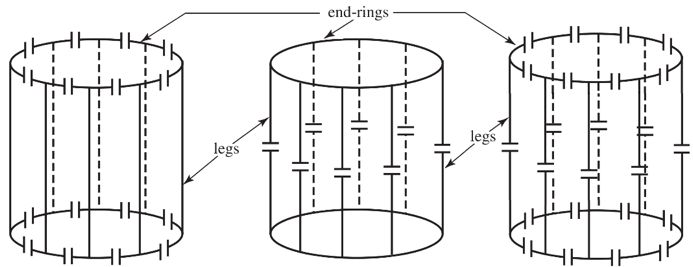}
	\caption{Birdcage Antenna: High-pass (left), Low-pass (middle), Band-pass (right), adapted from~\cite{739187}.}
	\label{fig:birdcages}
\end{figure}

\begin{figure}[h]
	\centering
	\includegraphics[width=7.7cm]{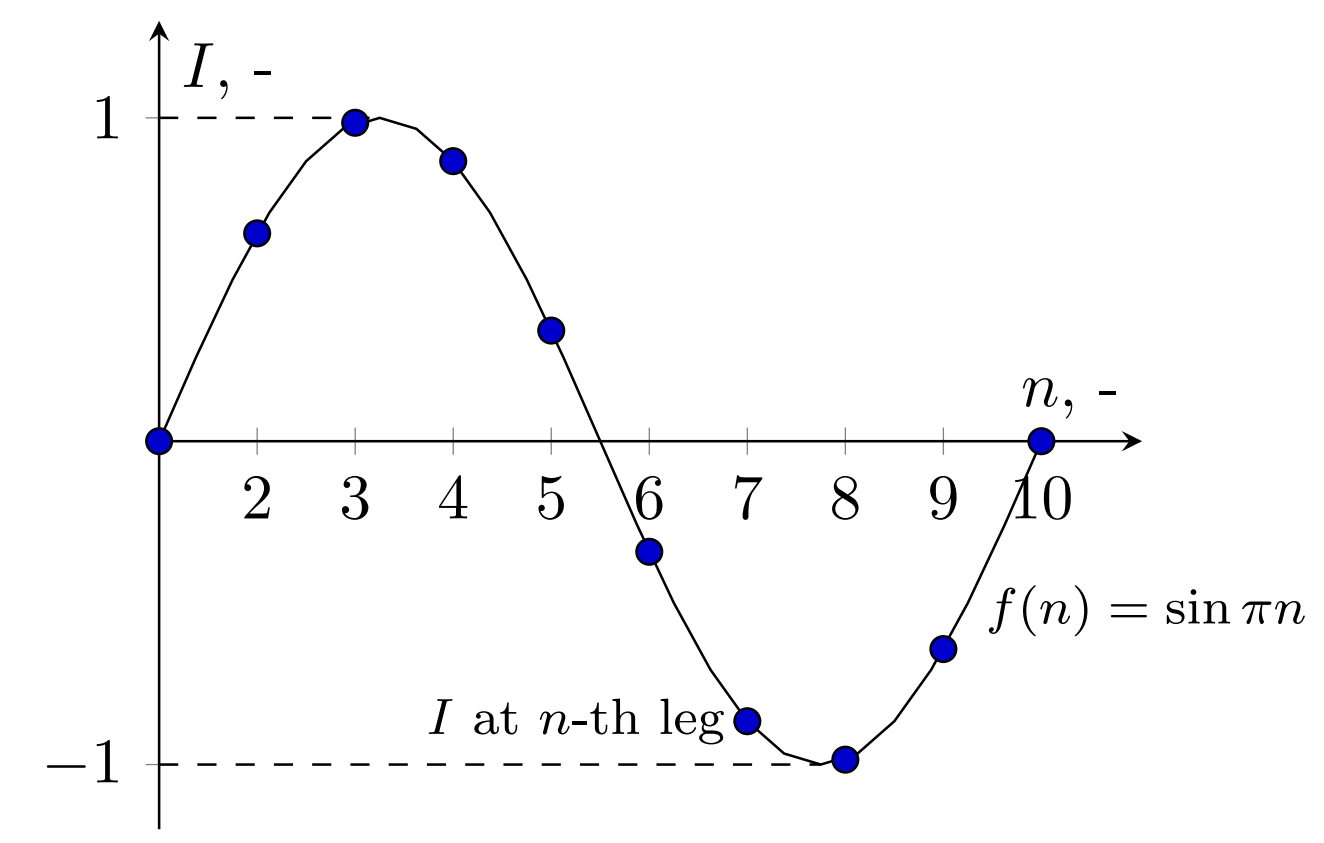}
	\caption{Current Distribution along a 10-leg Birdcage at a given Time.}
	\label{fig:birdcagecurrentI}
\end{figure}

\begin{figure}[h]
	\centering
	\includegraphics[width=7cm]{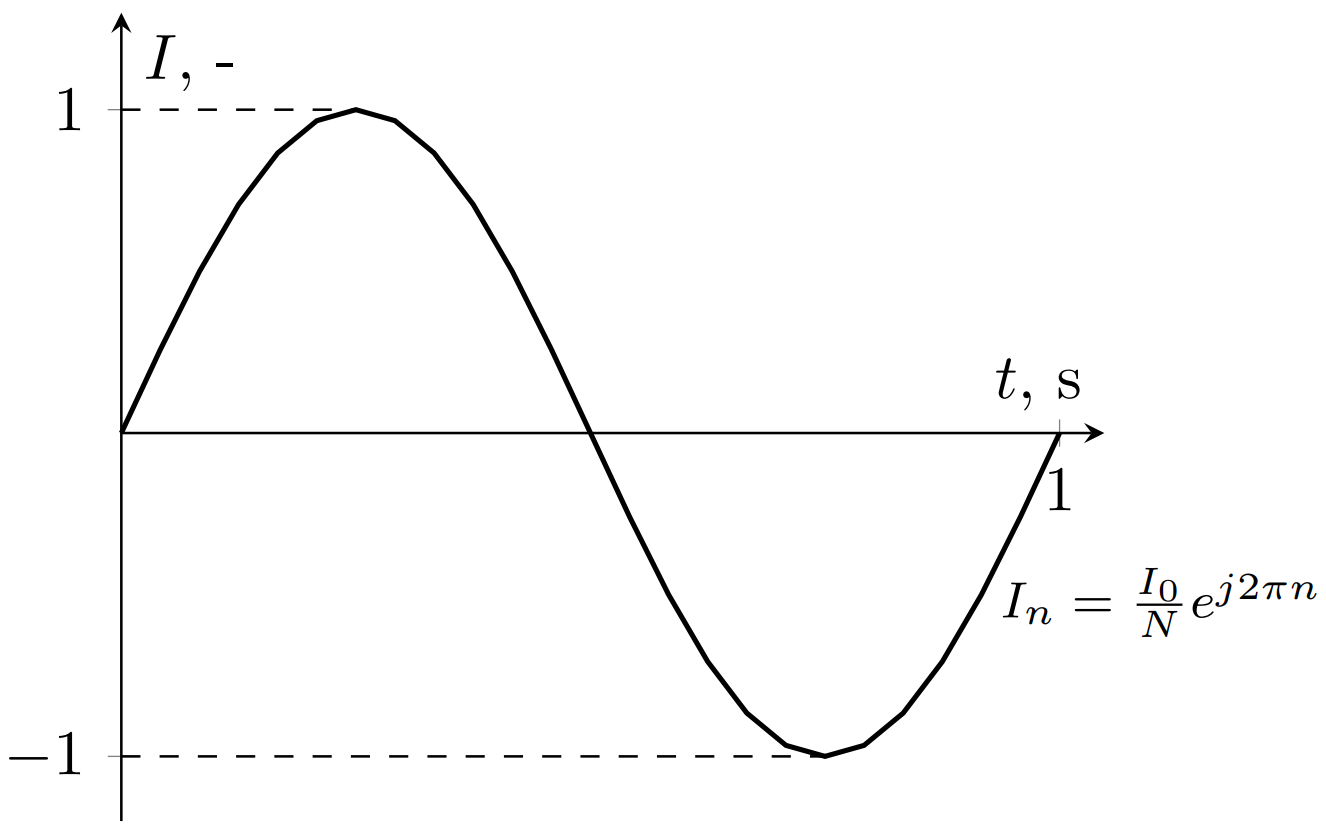}
	\caption{Current Amplitude over Time on a Single Birdcage Antenna Leg.}
	\label{fig:birdcagecurrentII}
\end{figure}

\begin{figure}[h]
		\centering
		\includegraphics[width=10cm]{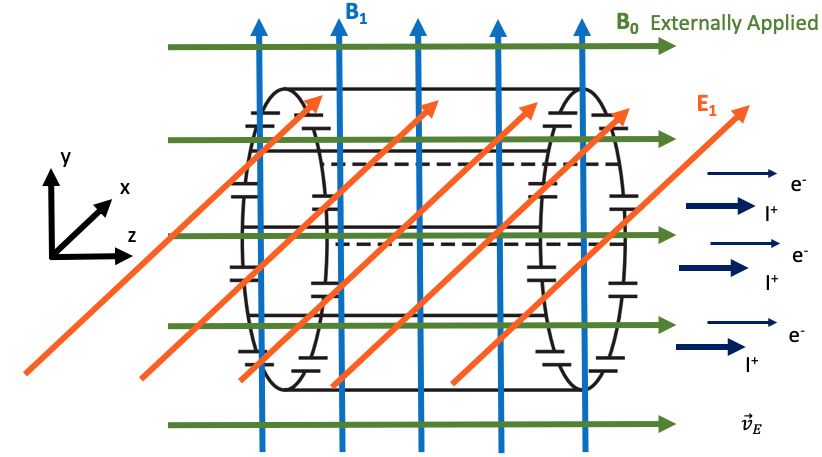}
		\caption{Birdcage EM fields $\vec{B}_1$,$\vec{E}_1$, Externally Applied Magnetic Field $\vec{B}_0$ (no plasma case).}
		\label{fig:bfieldconf}
	\end{figure}

Hereby, the theoretical description of the birdcage antenna is presented highlighting the advantages for the application in an RF helicon-based plasma thruster, in particular the operation at resonance condition and the EM fields configuration for providing partial acceleration of a quasi-neutral plasma plume.
The birdcage antenna has been originally developed for Magnetic Resonance Imaging (MRI)~\cite{birdcage1985,PASCONE1991395}. Birdcage antennae operate on the principle that a sinusoidal current distribution over a cylindrical surface induces a homogeneous transversal magnetic field within the cylindrical volume itself. Such antennae are made by two end-rings, connected by equally spaced legs with capacitors in between, and can be designed as low-pass, high-pass, or band-pass frequency response depending on the capacitors location, see Fig.~\ref{fig:birdcages}. The legs and/or the end-rings have capacitors in between to adjust the birdcage antenna resonance frequency to the one required by the application. 
The resonance frequency is that at which the impedance is totally real: the reactance $X$ is zero and its impedance $Z=R$. Therefore, the load is already partially matched and only the resistance $R$ requires further matching. Each birdcage antenna with $N$ legs has a spectrum of generally $k=N/2$ frequencies at which it resonates, called resonance modes. 
For the IPT, the high-pass mode is selected due to easier technical implementation. By feeding an high-pass birdcage antenna at one point at the resonance mode $k=1$, a linearly polarized homogeneous transversal magnetic field (usually required for MRI application) is provided~\cite{birdcage1985}. Consequently, also the electric field is linearly polarized, homogeneous, and perpendicular to the magnetic one within the cross section of the enclosed cylindrical volume. If the feeding is at two points quadrature, the EM fields are homogeneous and circularly polarized~\cite{birdcage1985}. The current distribution along the antenna follows the law described in Eq.~\ref{eq:current_birdcage}, where $I_{jk}$ is the normalized current at the $j$-th loop for the $k$-th mode of a birdcage antenna.\begin{equation}
\label{eq:current_birdcage}
I_{jk} = 
\begin{cases}
\cos{(\frac{2\pi j k}{N})};~k=0,1,2,...,N/2\\ \sin{(\frac{2\pi j k}{N})};~k=1,2,...,(N/2-1)
\end{cases}
\end{equation}
Therefore, the more legs are implemented, the better the current distribution matches a sinusoidal curve, see Fig.~\ref{fig:birdcagecurrentI}. On a single leg, instead, the amplitude of current variates over time as shown in Fig.~\ref{fig:birdcagecurrentII}.
Birdcage antennae are modelled by self and mutual inductances of legs and end-rings (ER) $L_{Leg}$ and $L_{ER}$, plus the applied capacitors of capacitance $C$. Resonance frequencies for the high-pass design are given in Eq.~\ref{eq:resonance}. The high-pass design has one resonance mode more ($k = 0$) at the highest frequency called anti-resonant AR.
	\begin{equation}
	\omega_{k_{HP}} = \biggl[C\biggl(L_{ER}+2 L_{Leg} \sin^2{\frac{\pi k}{N}}\biggr)\biggr]^{-1/2},~(k=0,1,2,...,N/2)
	\label{eq:resonance}
	\end{equation}
Only the resonant mode at $k=1$ presents the desired EM fields configuration. Those are shown in Fig.~\ref{fig:bfieldconf}. The magnetic field created by the birdcage is $\vec{B}_1$ along $y$, the respective electric field $\vec{E}_1$ is perpendicular to $\vec{B}_1$, along $x$. Since $\vec{B}_1$ is linearly polarized, its direction will switch along $y$ on each cycle, and, therefore, so will $\vec{E}_1$ along $x$~\cite{birdcage1985}. An additional external static magnetic field is provided along the $z$ axis $\vec{B}_0$ to aid the formation of helicon waves.
The resulting EM fields, which are due to the antenna only and still exclude those of the plasma, are expected to provide a drift velocity $\vec{v}_E=\vec{E}\times\vec{B}$ to both ions and electrons along the same direction on $z$. Correspondingly, thrust and exhaust velocity of significance are expected to be provided by this mode, indeed, such promising acceleration mechanism has been recently analytically investigated in~\cite{fruchtman_ExB}, while also ensuring a quasi-neutral plasma plume that does require the implementation of a neutralizer, see Eq.~\ref{eq:vdrift}. Such results support the use of a birdcage antenna for a contact-less plasma thruster application without needing a neutralizer. The $y$-component of the drift, instead, is zeroed at each cycle due to the linear polarization of the EM fields produced by the birdcage antenna. In the exhaust region, the applied magnetic field $\vec{B_0}$ is diverging, therefore partial thrust will also be provided by an EM nozzle effect, this can be investigated by numerical methods such as~\cite{magarotto20203d,gallina2019enhanced,numeric}, however this is out of scope within this paper. 
\begin{equation}
\vec{v}_E = \frac{1}{\vec{B}^2} \begin{Vmatrix}
\hat{x} & \hat{y} & \hat{z} \\
E_1 & 0 & 0 \\
0 & B_1 & B_0 \\
\end{Vmatrix} = \frac{1}{{B_0}^2+{B_1}^2}\begin{Bmatrix} 0 \\ -E_1 B_0 \\ E_1 B_1 \end{Bmatrix}
\label{eq:vdrift}
\end{equation}

\section{IPT Design with a Birdcage Antenna}
\begin{figure}[h]
	\centering
	\includegraphics[width=7cm]{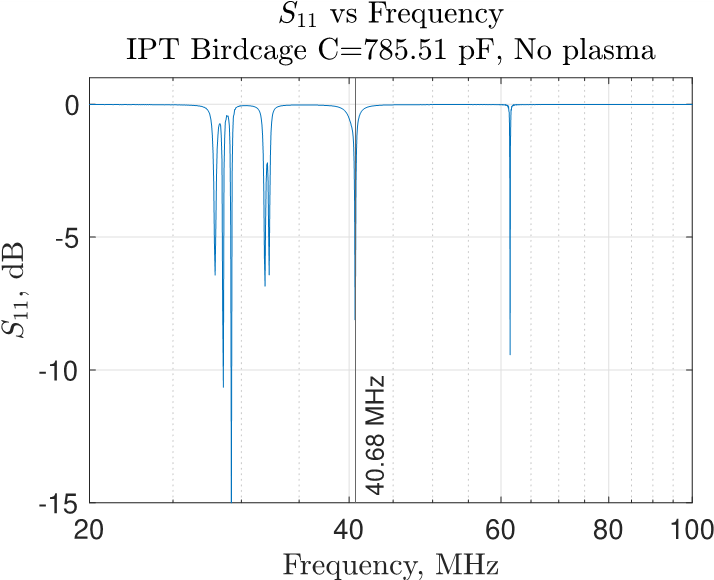}
	\caption{$S_{11}$ vs Frequency IPT.}
	\label{fig:S11}
\end{figure}

\begin{figure}[h]
	\centering
	\includegraphics[width=7cm]{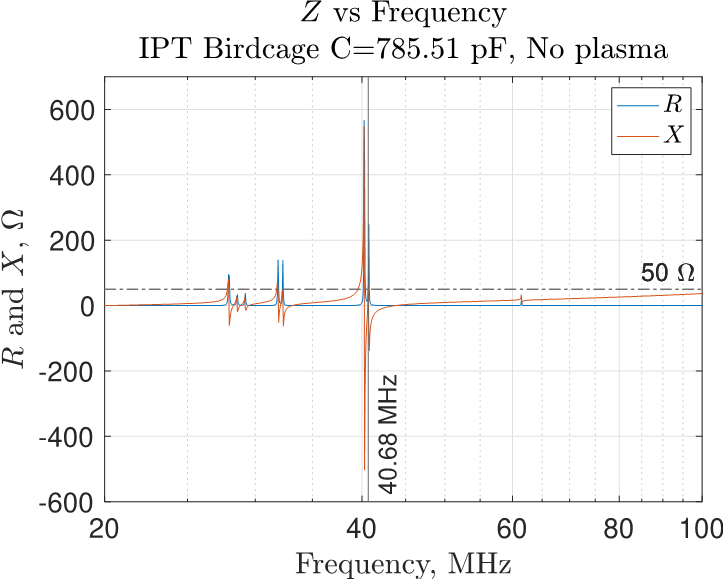}
	\caption{$Z$ vs Frequency IPT.}
	\label{fig:Z}
\end{figure}

\begin{figure}[h]
	\centering
	\includegraphics[width=11cm]{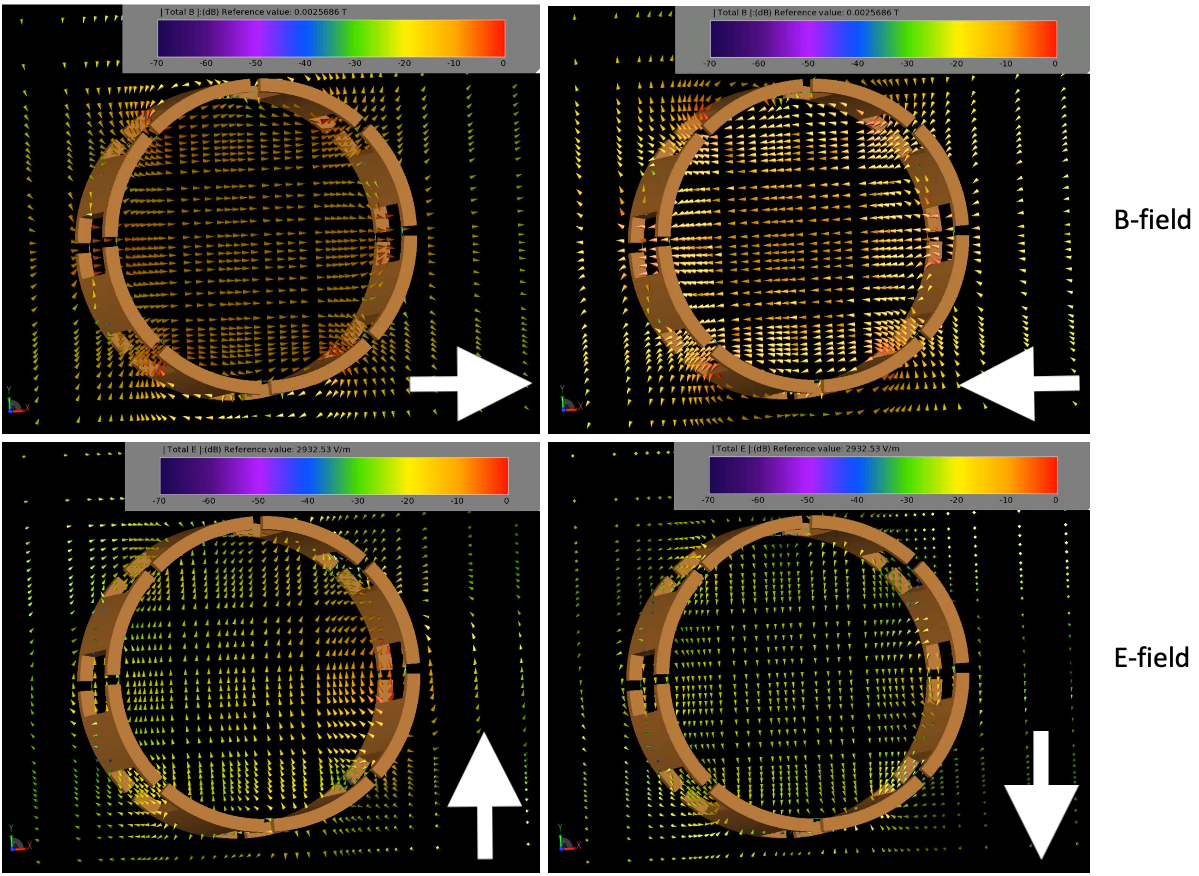}
	\caption{Magnetic Field (above) and Electric Field (below) during one Cycle for a Birdcage Antenna: Linear Polarization~\cite{romanoiepc3}.}
	\label{fig:EMF2}
\end{figure}

 A birdcage with $8$ legs in a high pass design has been chosen and designed for resonance at \SI{40.68}{\mega\hertz} at the $k=1$ mode, feeding is provided at one point for linear polarization of the EM fields. The commercial 3D EM simulation software Remcom Inc. XFdtd\textsuperscript{\textregistered} 7.8.1.3 is used to evaluate the resonance spectrum, the corresponding impedances, the EM fields visualization, and the resulting required capacitance. In XFdtd\textsuperscript{\textregistered}, the frequency is swept as signal input to the antenna, and the resulting impedance $Z$ and scattering parameter $S_{11}$ over frequency are extracted. The $S_{11}$ parameter is generally defined as the input port voltage reflection coefficient~\cite{balanis}. It represents how much power is absorbed and reflected by a load depending on the input at the signal source. The resonance study is shown for $S_{11}$ and for $Z$ in Fig.~\ref{fig:S11} and Fig.~\ref{fig:Z} from XFdtd\textsuperscript{\textregistered}. The estimated required capacitance to match the desired resonance frequency mode is of $C=\SI{785.51}{\pico\farad}$. Each peak of $S_{11}$ is a resonance mode, and, correspondingly, the reactance goes to $X=\SI{0}{\ohm}$. Indeed, there are $N/2=4$ resonance modes plus the AR one at the highest frequency. To verify that the $k=1$ is the correct one, the 3D EM fields are extracted and visualized over one RF cycle in the cylindrical cross section enclosed by the birdcage, see Fig.~\ref{fig:EMF2}. Those have been compared to those of  the other resonance modes.
An example of the correct resonance with a linearly polarized field is shown in Fig.~\ref{fig:EMF2} (top) for the magnetic field and in Fig.~\ref{fig:EMF2} (bottom) for the electric field~\cite{romanoiepc3}. At each RF cycle the EM fields reverse their direction due to the linear polarization. Once the correct resonance peak is identified, the capacitance is swept until the targeted peak is at \SI{40.68}{\mega\hertz}. The externally applied magnetic field is expected to trigger the formation of helicon waves within the discharge channel therefore providing a higher plasma density and a more efficient ionization~\cite{chen224,EPFL4}.
As can be seen, the resonance frequency of $k=1$ presents $S_{11}=\SI{-8.4}{\decibel}$ that translates in the fact that the real part $R$ of the impedance $Z$ is not perfectly matched to that of the RF generator output, and that requires further matching/tuning of $R$ only to reach a more desired $S_{11}<\SI{-20}{\decibel}$ for having $>99\%$ power coupling.

\subsection{Resonance Tuning}
The design for resonance ensures that such condition is maintained without plasma. Once the plasma is ignited, a frequency shift and a corresponding impedance change will happen. This requires tuning to shift back to the original resonance condition~\cite{EPFL1}. Such effect has been already seen and solved in MRI machines, as it is caused by the introduction of parts of bodies within the antenna~\cite{HAYES1985622,ozen2017novel}. Tuning can be realized, by moving conducting components that change $X_C$ and $X_L$ next to the antenna~\cite{HAYES1985622,ozen2017novel}. This is implemented in the IPT by the movable injector made of conductive material (brass), preliminary tests showed the capability to shift up to $+\SI{1}{\mega\hertz}$. Moreover it is expected that tuning can be performed (with plasma present) also by changing position and strength of the applied static magnetic field as reported in~\cite{EPFL1}. The preliminary resonance verification is shown within Fig.~\ref{fig:comparison}, ignition is expected by early 2020.

\subsection{Resonance Verification}
\begin{figure}[h]
	\centering
	\includegraphics[width=10cm]{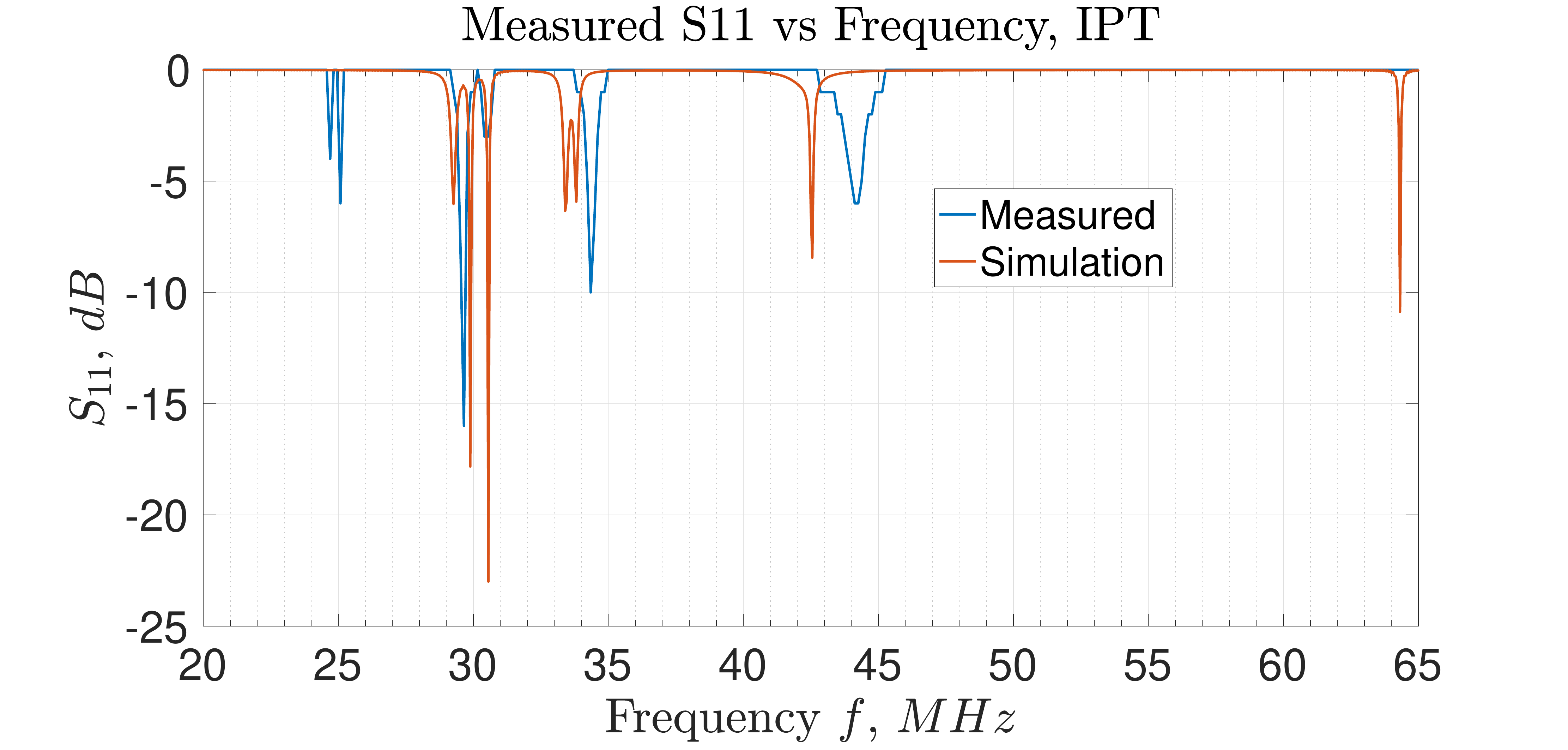}
	\caption{$S_{11}$ vs $f$ Simulation and Measurement IPT, $C=\SI{717.475}{\pico\farad}$.}
	\label{fig:comparison}
\end{figure}

The IPT has been finally assembled and tested with a spectrum analyzer, FSH4 from Rohde \& Schwarz that operates in the range of \SI{100}{\kilo\hertz} to \SI{3.6}{\giga\hertz}. 
The finally achievable capacitance is due to the availability in the capacitors market, resulting in an average of $C=\SI{717.475}{\pico\farad}$. Results show a good agreement with the simulations performed with XFdtd\textsuperscript{\textregistered}, see Fig.~\ref{fig:comparison}. The relative difference in terms of frequency for the desired peak of interest $k=1$ is of $3.84\%$, \SI{42.55}{\mega\hertz} estimated vs \SI{44.25}{\mega\hertz} measured. The $S_{11}$ according to simulation is of $\SI{-8.4}{\decibel}$ while for the measured one is of $\SI{-6.0}{\decibel}$. Both differences are acceptable as there are too many uncertainties and the final antenna adjustment will be performed manually plus, a mechanism of fine tuning is included into the IPT design. In particular, such differences are due a combination of the following aspects:\begin{itemize}
\item Capacitors are considered ideal, zero inductance, in the XFdtd\textsuperscript{\textregistered} simulations;
\item XFdtd\textsuperscript{\textregistered} simulations take into account an average value for the capacitance to be equal for all capacitors;
\item Integration of the birdcage is not ideal $\Rightarrow$ tolerances in the birdcage legs manufacturing, their alignment, as well as their contact surfaces are different.
\end{itemize}
Once the resonance frequency has been fixed, the relative position between RF input power and ground on the antenna can be optimized to achieve the best coupling in terms of respective impedance matching~\cite{EPFL5}.

\section{Conclusion and Outlook}
The IPT based on birdcage antenna has been designed, simulated, assembled, and preliminary verified. The birdcage antenna operating at resonance ensure a partially matched load by having $X=\SI{0}{\ohm}$. Moreover, the IPT EM fields configuration is expected to produce a $\vec{E}\times\vec{B}$ drift of ions and electrons towards the same direction, combined with an EM nozzle effect, thus providing thrust and a (plasma thrusters typical) quasi-neutral plasma plume, finally removing the need of a neutralizer. The movable conductive injector is planned to be used for tuning the resonance frequency, as well as the externally applied static magnetic field~\cite{HAYES1985622,ozen2017novel,EPFL1}. Verification of the XFdtd\textsuperscript{\textregistered} simulations against spectrum analyzer show good agreement, however final tuning always need to be performed. By early 2020 the IPT will be attempted a first ignition followed by a discharge characterization for different gases, mass flows, powers, and magnetic field strengths. Afterwards, the plasma will be evaluated by respective diagnostic tools such as Langmuir and Faraday probes at first, Retarding Potential Analyzer and optical emission spectroscopy (OES) as next.

\section*{Acknowledgements}
This project has received funding from the European Union's Horizon 2020 research and innovation programme under grant agreement No.~737183. This reflects only the author's view and the European Commission is not responsible for any use that may be made of the information it contains.

\newpage
\bibliographystyle{elsarticle-num}
\bibliography{bibliography}

\end{document}